\documentclass[reprint,aps,pra,nofootinbib,superscriptaddress]{revtex4-1}
\usepackage{xcolor,graphicx}
\usepackage{amssymb,amsmath,amsthm,mathtools}
\usepackage{verbatim,float}
\usepackage{bbm,multirow}
\usepackage[thinlines]{easytable}
\usepackage{braket}
\usepackage[varg]{txfonts}
\usepackage{hyperref}
\usepackage{array}
\usepackage{bbding}
\usepackage{pifont}
\usepackage{wasysym}

\newcommand{\tx}{\text} 
\newcommand{\mf}{\mathfrak}
\newcommand{\mc}{\mathcal}

\newcommand{\mbb}{\mathbbm}

\hypersetup{colorlinks=true,linkcolor=black,urlcolor=blue,citecolor=blue}

\DeclareMathOperator{\Tr}{\mathrm{Tr}}

\begin{document}

\title{Genuine multipartite entanglement and quantum coherence in an electron-positron system: Relativistic covariance}

\author{Alexandra T. Petreca}
\affiliation{Instituto de F\'isica, Universidade de S\~ao Paulo, Caixa Postal 66318, CEP 05315-970 S\~ao Paulo, S\~ao Paulo, Brazil}

\author{Gabriel Cardoso}
\affiliation{Department of Physics and Astronomy, Stony Brook University, Stony Brook, New York 11794, USA}

\author{Fernando P. Devecchi}
\affiliation{Departamento de F\'isica, Universidade Federal do Paran\'a, PO Box 19044, 81531-980 Curitiba, Paran\'a, Brazil}

\author{Renato M. Angelo}
\affiliation{Departamento de F\'isica, Universidade Federal do Paran\'a, PO Box 19044, 81531-980 Curitiba, Paran\'a, Brazil}

%===============
\begin{abstract}
The last two decades have witnessed an increasing effort by the scientific community toward pursuing a better framework for quantum resource covariance, with the focus predominantly posed on quantum entanglement. In this work, we move the discussion one step further by analyzing the behavior of both genuine multipartite entanglement and quantum coherence under Lorentz boosts. Specifically, we conduct a case study for the problem of an electron-positron pair created in a superposed multipartite pure state. Our approach is different from the standard treatments also in that we consider all the components of the four-momentum, thus allowing for an inspection of scenarios wherein  entanglement can be encoded among these degrees of freedom as well. Our analysis reveals interesting subtleties in this problem, such as the fact that genuine 4-partite entanglement in the laboratory frame transforms into genuine 8-partite entanglement plus quantum coherence in the perspective of the Lorentz-boosted frame. Moreover, a given combination of these quantum resources is shown to form a Lorentz invariant. Although our findings are not able to determine, via first principles, an information-theoretic Lorentz invariant, they pave the way for fundamental incursions along this line.
\end{abstract}
%===============
\maketitle

%%%%%%%%%%%%%%%%%%%%%%%%%
\section{Introduction}
\label{sec:introduction}
%%%%%%%%%%%%%%%%%%%%%%%%%

Several decades have passed since seminal works brought to light the quantum mechanical phenomenon of entanglement \cite{einstein1935can,bell1964einstein}, whose conceptual and practical implications have consistently challenged physicists to this day. Entanglement is by now widely acknowledged as a quantum resource \cite{horodecki2009} for tasks such as quantum cryptography \cite{gisin2002}, randomness generation \cite{pironio2010}, and quantum metrology \cite{toth2014}. Many other quantum resources have since been discovered, among which quantum coherence \cite{QCoh2014} emerges as a critical one for many operational tasks \cite{marvian2016}. 

In light of the fundamental relevance of quantum resources to both foundational and technological developments, it is natural to ask how they behave as the systems are allowed to enter the high-velocity domain and how much of these resources are available to different observers. Most importantly, although violations of the no-signaling principle have never been observed, the strict relation found between pure-state entanglement and Bell nonlocality \cite{brunner2014} has prompted researchers to inspect the frontier between quantum mechanics and special relativity. In effect, over the past two decades, numerous works have tested the relativistic invariance of entanglement on a number of physical systems.

With the goal of studying entanglement and Bell nonlocality in states of relativistic spins, the standard strategy consists of starting with free particle states encoding momentum and spin degrees of freedom, and then tracing over momentum  \cite{gingrich2002quantum,alsing2002lorentz,li2003relativistic,lee2004quantum,caban2005covariant,jordan2007lorentz,chakrabarti2009entangled,friis2010relativistic,choi2011diracparticles,rastgoo2014spin,palge2015behavior,palge2018relativistic,fan2018relativistic,bittencourt2018effects,bittencourt2018global}. In this scenario, a variety of interesting questions arises, including investigations regarding the operational significance of spin \cite{lee2004quantum,czachor2005comment,caban2005covariant,choi2011diracparticles,saldanha2012physical,giacomini2019relativistic}. A general conclusion about entanglement between relativistic particles is that it is invariant for bipartitions which keep the momentum and spin degrees of freedom of each particle together. This derives from the fact that a Lorentz boost $U(\Lambda)$ for a two-particle system is given by the product $U_1(\Lambda)\otimes U_2(\Lambda)$, which, being local, cannot change the entanglement between the particles. On the other hand, $U_k(\Lambda)$ ($k=1,2$) does change the amount of entanglement between the spin and the momentum of the $k$ th particle \cite{peres2002quantum,caban2005covariant,palge2009,rastgoo2014spin,bittencourt2020single}. Further discussions on this line of research include the differences in treating  momentum as a discrete \cite{friis2010relativistic,palge2015behavior} or continuous \cite{rastgoo2014spin,palge2018relativistic} variable. Bell inequality violations \cite{lee2004quantum}, scattering states \cite{fan2018relativistic}, entanglement between spin and parity eigenstates (for spin-$1/2$ fermions) \cite{bittencourt2018global,bittencourt2020single,bernadini2020lorentz}, and the inclusion of non-inertial and gravitational effects  \cite{friis2012motion,alsing2012observer}. The reader is also referred to Refs. \cite{li2003relativistic,peres2004quantum,moon2004relativistic,bartllet2005relativistically,he2007quantum,saldanha2012physical,giacomini2019relativistic} for the consequences that the covariance properties of entanglement have for relativistic quantum information science.

Whenever a system of two or more particles is concerned, the characterization of entanglement becomes subtler. This is so because the existence of more than two Hilbert spaces yields tricky notions such as multipartite entanglement \cite{thapliyal1999} and genuine multipartite entanglement \cite{loock2003,ma2011measure}. Research regarding the manifestation of such resources for relativistic particles is still incipient \cite{huber2011lorentz,amiri2014}. In addition, the typical approach leaves out the fact that relativistic momentum actually needs four Hilbert spaces to be properly described in terms of its components, among which entanglement can, in principle, be generated. Another important gap in the field of relativistic quantum information is the lack of systematic studies on quantum resources other than entanglement and their eventual interconnections upon changes of reference frames. Interestingly, results along these lines have recently been reported in the arena of quantum reference frames: although two ``quantum observers'' do not agree on the amounts of entanglement and quantum coherence \cite{giacomini2019}, they always agree upon the total amount of quantum resources, as long as quantum incompatibility is included in the description \cite{savi2021}.

This work is aimed at contributing to the above discussion. Specifically, we want to make the point that even though genuine multipartite pure-state entanglement and quantum coherence are not invariant under Lorentz boosts, a given combination of these resources turns out to be. To accomplish this task, we proceed as follows. The starting point for deriving the relativistic particle states and their transformation properties is the relativistic wave equations or, as an extension, the corresponding quantum fields \cite{alsing2002lorentz,soo2004vonneumann,choi2011diracparticles,bittencourt2018effects,bittencourt2018global}. As is now well known,  covariance aspects can thus be treated in a way convenient for quantum mechanics problems, using concepts such as Wigner rotations on single-particle states. We apply this formalism to a specific problem: evaluating the quantum resources encoded in a given quantum state of a particle-antiparticle system resulting from a pair-production event \cite{alsing2002lorentz,florio2021gibbs} from the perspectives of the laboratory and a boosted frame. After computing entanglement for every bipartition and quantum coherence for every reduced state as a function of the parameters of the quantum state and the boost, we evaluate genuine multipartite entanglement and come up with an information-theoretic Lorentz invariant. 

%%%%%%%%%%%%%%%%%%%%%%%%%%%%%%
\section{Preliminary concepts}
\label{sec:prelimiaries}

%%%%%%%%%%%%%%%%%%%%%%%%%%%%%%
\subsection{Wigner Rotations}
\label{section:wigner_rotations}

To conveniently represent the action of Lorentz transformations on each single-particle state, we follow the prescription reviewed in \cite{alsing2002lorentz,halpern1968special}. First, one expresses each state $\Psi_{p,\sigma}$ of momentum $p=(p_0,\vec{p})$ and spin $\sigma$ through a Lorentz transformation $L(p)$ from a standard inertial reference frame, in which the particle state reads $\Psi_{k,\sigma'}$, with momentum $k$ and spin $\sigma'$. Formally,
\begin{equation}
\Psi_{p,\sigma}=N(p)U\left(L(p)\right)\Psi_{k,\sigma'},
\label{eq:stdstate}
\end{equation}
where $N(p)=\sqrt{k_0/p_0}$ is a constant of normalization, which we will drop in the end,\footnote{This constant follows from quantum field theory conventions, where it comes in to guarantee that the measure of integration in momentum space is Lorentz invariant. Since we treat momentum as a discrete observable, this constant will be absorbed in the normalization of its eigenstates.} and $U\left(L(p)\right)$ is the $4\times 4$ matrix representing the Lorentz transformation in the Dirac spinor representation. Here we take the preferred inertial reference frame to be the particle's rest frame, in which $k=(mc,0,0,0)$. The application of a Lorentz transformation $\Lambda$ to $\Psi_{p,\sigma}$ results in a state with momentum $\Lambda p$, namely, 
\begin{equation}
U(\Lambda)\Psi_{p,\sigma}=\sqrt{\frac{(\Lambda p)_0}{p_0}}\sum_{\sigma'} D_{\sigma \sigma'} \left(W(\Lambda,p)\right)\Psi_{\Lambda p,\sigma'},
\end{equation}
with $D(W(\Lambda,p))$ the representation of the \emph{Wigner rotation}\footnote{That $W(\Lambda,p)$ is a rotation matrix follows from the fact that it is in the subgroup of the Lorentz group which keeps the momentum $k$ fixed.} $W(\Lambda,p)=L^{-1}(\Lambda p)\Lambda L(p)$ in the spin basis. It is a well-known fact about the Lorentz group $\mathrm{O}(1,3)$ that the transformation $W$ lies in the Little Group $\mathrm{SO}(3)\subset \mathrm{O}(1,3)$.

\begin{figure}[htb]
\centering
\includegraphics[scale=0.35]{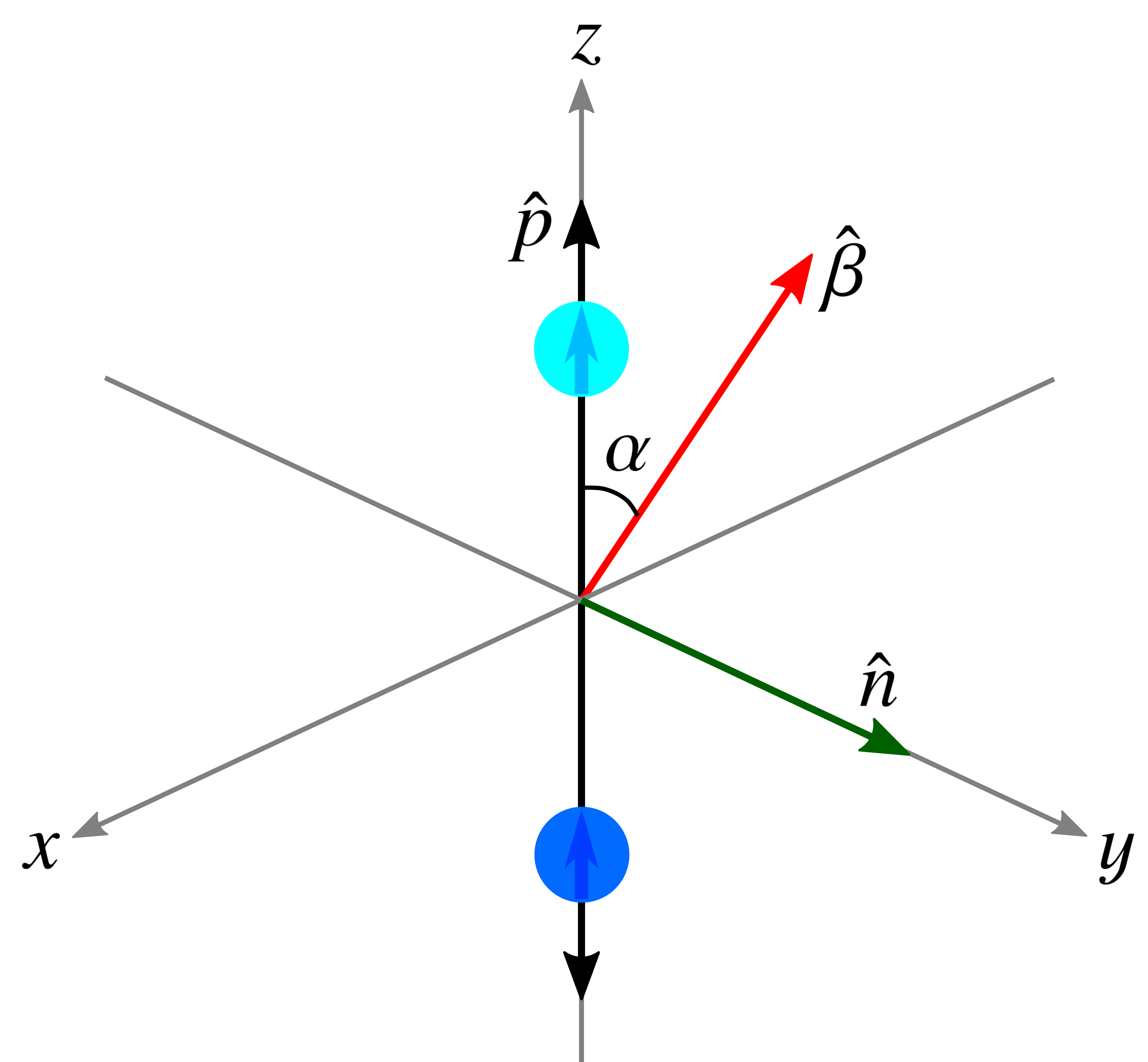}
\caption{A spin-$\frac{1}{2}$ particle (cyan sphere) and its antiparticle (blue sphere) move along the directions defined by the momenta unit vectors $\hat{p}=\hat{z}$ and $-\hat{p}$, respectively, both with ``spin up'' $\ket{0}$, in the computational basis. A boost is performed in the $\hat{\beta}=\hat{z}\,\cos\alpha-\hat{x}\,\sin\alpha$ direction. The unit vector $\hat{n}=\hat{\beta}\times\hat{p}/||\hat{\beta}\times\hat{p}||$ defines the axis around which the Wigner rotation is performed for the particle, whereas for the antiparticle, the respective axis is given by $-\hat{n}$.}
\label{fig:setup}
\end{figure}

Let us write the Wigner rotation for a massive spin-$\frac{1}{2}$ (Dirac) particle \cite{alsing2002lorentz} moving along the $z$ direction with momentum $\vec{p}=\mf{p}\hat{z}$, where $\mf{p}=\gamma_vmv$, $\gamma_v=(1-\beta_v^2)^{-1/2}$, and $\beta_v=v/c$. This instance is depicted in the upper half of Fig.~\ref{fig:setup}, which anticipates the physical setup we shall focus on throughout this work. The boost is taken to lie in the $xz$ plane, in the direction defined by the unit vector $\hat{\beta}=\hat{z}\,\cos\alpha-\hat{x}\,\sin\alpha$, which is parametrized by an angle $\alpha\in[0,\pi]$ measured from the positive $z$ axis in the direction of the negative $x$ axis, and with the Lorentz factor $\gamma=(1-\beta^2)^{-1/2}$. It follows that $W(\Lambda,p)$ is a rotation (around the $y$ axis) with matrix representation
\begin{equation}
W(\Lambda,p)=
\begin{pmatrix}
1 & 0 & 0 & 0 \\
0 & \cos\Omega(\Lambda,p) & 0 & \sin\Omega(\Lambda,p)\\
0 & 0 & 1 & 0\\
0 & -\sin\Omega(\Lambda,p) & 0 & \cos\Omega(\Lambda,p)
\end{pmatrix},\label{W}
\end{equation}
where the rotation angle is given by \cite{halpern1968special}
\begin{equation}
\cos\Omega=\frac{\gamma+\gamma_v+\beta\beta_v\gamma\gamma_v\cos\alpha+(1-\gamma-\gamma_v+\gamma\gamma_v)\cos^2\alpha}{1+\gamma\gamma_v+\beta\beta_v\gamma\gamma_v\cos\alpha}.
\label{cos-Omega}
\end{equation}

It is instructive to look at a few interesting particular cases. For a Lorentz boost in the $z$ direction, the Wigner rotation is trivial since, in this case, $W(\Lambda,p)=L^{-1}(\Lambda p) L(\Lambda p)=I_4$. Indeed, with $\alpha=0$, we find $\cos\Omega(\Lambda,p)=1$, which leads Eq. \eqref{W} to the mentioned result. Conversely, in the case of a boost along the $-x$ direction ($\alpha=\pi/2$), we get the known result $\cos{\Omega}=(\gamma+\gamma_v)/(1+\gamma\gamma_v)$ \cite{alsing2002lorentz,palge2009}. Other results can be readily checked from the general formula \eqref{cos-Omega}: (i) for any $\alpha$, the small-velocity limit (either $\beta=0$ or $\beta_v=0$) recovers the identity, $W=I_4$, and (ii) in the relativistic limit (both $\beta\to 1$ and $\beta_v\to 1$), the Wigner angle is simply given by $\Omega\to\alpha$.

The four-dimensional space of Dirac spinors can be decomposed into two-dimensional particle and antiparticle subspaces as $\mathcal{H}=\mathcal{H}^+\oplus\mathcal{H}^-$. Let us introduce the (computational) basis of the spin-$z$ eigenstates in each space $\mathcal{H}^{\nu}$ through the relation $\sigma_z\ket{s}^{\nu}=(-1)^s\ket{s}^{\nu}$, with  $s\in\{0,1\}$ and $\nu\in\{-,+\}$ (referring to electron and positron, respectively). In this basis, the Wigner rotation is represented by \cite{tong2006lectures}
\begin{equation}
D(W(\Lambda,p))=\mf{D}_{\hat{n}}(\Omega)\oplus \mf{D}_{\hat{n}}(\Omega),
\end{equation}
where $\mf{D}_{\hat{n}}(\Omega)\coloneqq e^{-i(\Omega/2)\hat{n}\cdot\vec{\sigma}}$, with $\vec{\sigma}$ the vector formed by the Pauli matrices, and $\hat{n}=\hat{\beta}\times\hat{p}/||\hat{\beta}\times\hat{p}||$. Thus, for one-particle states, a simple transformation law follows, namely, $U(\Lambda)\ket{p,s}^\nu = \ket{\Lambda p}^\nu \mf{D}_{\hat{n}}(\Omega)\ket{s}^\nu$. Specializing for the scenario of our interest (Fig. \ref{fig:setup}), now using the notation $p^\mf{s}=(\mf{p}_0,0,0,\mf{sp})$ and  $\Lambda p^\mf{s}=\pi^\mf{s}$, we can write
\begin{equation}
U(\Lambda)\ket{\mf{p}_0,0,0,\mf{s}\mf{p},s}^\nu = \ket{\pi_0^\mf{s},\pi_x^\mf{s},0,\pi_z^\mf{s}}^\nu \mf{D}_{\mf{s}\hat{y}}(\Omega)\ket{s}^\nu,
\label{Uket}
\end{equation}
where $\mf{s}\in\{-,+\}$ is the sign of the $z$ component of momentum. As can be checked by direct application of the Lorentz transformations, the new four-momentum $\pi^\mf{s}$ implicitly depends on $\mf{s}$ in a way such that $\pi_0^+=\pi_0^-$ iff $\alpha=\frac{\pi}{2}$ and $\pi_x^+=\pi_x^-$ iff $\alpha\in\{0,\frac{\pi}{2},\pi\}$, whereas $\pi_z^+$ never equals $\pi_z^-$. Hence, by restricting the boost to the domain $\alpha\in\big(0,\frac{\pi}{2}\big)$, we ensure that $\pi^+\neq \pi^-$. As a consequence, we have $\braket{\pi_a^\mf{s}|\pi_a^\mf{s'}}=\delta_{\mf{ss'}}$ for $a\in\{0,x,z\}$.

%%%%%%%%%%%%%%%%%%%%%%%%%%%%%%%%%%%
\subsection{Electron-positron pair}
\label{sec:ep-pair}

Now we specify the arena wherein our study will be conducted. The system of interest is a positron-electron pair whose creation is chosen to occur on the $z$ axis (see Fig.~\ref{fig:setup}). Because the pair creation has to satisfy the conservation laws, one can always find an inertial reference frame in which the two particles move with opposite momenta $\pm\mf{p}\hat{z}$ and $\mp\mf{p}\hat{z}$, while the total energy is the same irrespective of the sign that is adopted. As far as the spin is concerned, the state of affairs prior to the matter creation, which basically involved a photon and a nucleus, demands that both the electron and positron manifest their spins in the same direction,\footnote{In the language of QED, this is understood as a consequence of the interaction Lagrangian (or, equivalently, the Feynman rules for the vertex diagram), which is proportional to $\epsilon_{\mu}\bar{v}_{\alpha}\gamma^{\mu}_{\alpha\beta}u_{\beta}$ and therefore imposes the conservation of angular momentum of the photon (fixed by the polarization state $\epsilon^{\mu}$) and those of the spinors (fixed by the spin states $\bar{v}_{\alpha}$, $u_{\beta}$).} being either ``spin up'' ($\ket{0}$) or ``spin down'' ($\ket{1}$). The situation depicted in Fig.~\ref{fig:setup} is then expressed as $\ket{\mf{p}_0,0,0,\mf{p},0}\ket{\mf{p}_0,0,0,-\mf{p},0}$, where we have dropped the system descriptor $\nu$ in favor of a notation that maintains the electron kets always on the left-hand side. Quantum mechanics also allows for the occurrence of a state such as $\ket{\mf{p}_0,0,0,-\mf{p},1}\ket{\mf{p}_0,0,0,\mf{p},1}$ in consonance with the same conservation laws and, more generally, a superposition of these two scenarios. Now, as can be seen in Eq. \eqref{Uket}, the parts of the state associated with the $y$ component of momentum will not change upon the considered boosts and, therefore, will remain uncorrelated with the remainder degrees of freedom. This means that the whole description can be done in terms of one-particle states $\ket{p_0,p_x,p_z,s}$. We then propose to describe the electron-positron state in the simplified form
\begin{equation}
\ket{\Psi} = \eta \ket{\mf{p}_0,0,\mf{p},0}\ket{\mf{p}_0,0,-\mf{p},0} + \xi e^{i \theta} \ket{\mf{p}_0,0,-\mf{p},1}\ket{\mf{p}_0,0,\mf{p},1},
\label{Psi_before}
\end{equation}
with amplitudes $\eta = \cos \phi$ and $\xi = \sin \phi$, with  $\phi \in \big[0,\frac{\pi}{2}\big]$, and a generic phase $\theta$. As usual, momentum is treated as a discrete variable, so that $\braket{\epsilon_1\mf{p}|\epsilon_2\mf{p}}=\delta_{\epsilon_1\epsilon_2}$ for $\epsilon_{1,2}\in\{+,-\}$. In fact, in both reference frames, each momentum component can be effectively treated as a qubit. 

%%%%%%%%%%%%%%%%%%%%%%%%%%%%%%%%%%%%%%%%%%%%%%
\subsection{Genuine multipartite entanglement}
\label{sec:GME}

The advancement of quantum computation and quantum cryptography urged the development of quantum resource {\it quantifiers} \cite{horodecki2009,QCoh2014}, a task that is by no means trivial, especially when multipartite states are concerned. Here, we restrict our attention to the quantification of genuine multipartite entanglement (GME) for pure states, a class of entanglement that refers to a ``global link'' not conceivable in terms of the correlations existing between bipartitions. To make this notion more rigorous, let us start with a system composed of three parts (degrees of freedom) denoted $1$, $2$, and $3$. A pure state $\ket{\psi}$ is said to be biseparable if $\ket{\psi}= \ket{\psi_j} \otimes \ket{\psi_{kl}}$, with $\{j, k, l\}$ assuming any permutation of $\{1,2,3\}$, totally separable if $\ket{\psi}= \ket{\psi_1} \otimes \ket{\psi_2} \otimes \ket{\psi_3}$, and genuinely entangled if it is neither biseparable nor totally separable. In the latter case, the state is said to possess genuine $3$-partite entanglement. Of course, the situation becomes more and more complex as one adds degrees of freedom to the system.

Let us now consider a $n$-partite state $\ket{\psi}\in \bigotimes_{i=1}^n\mc{H}_i$. In Ref.~\cite{ma2011measure}, a measure $\mbb{E}_n(\psi)$ has been put forward to quantify the amount of genuine $n$-partite entanglement encoded in the pure state $\psi=\ket{\psi}\bra{\psi}$ for an arbitrary $n$. Called generalized concurrence, this measure formally reads
\begin{equation}
\mbb{E}_n(\psi)\coloneqq\min_{\wp_i \in \mf{P}} \sqrt{2 \mc{L}(\rho_{\wp_i})},
\label{En}
\end{equation}
where $\mc{L}(\rho)\coloneqq 1-\Tr(\rho^2)$ is the linear entropy of $\rho$ (a measure of pure-state bipartite entanglement when $\rho$ is a reduced state) and $\mf{P}$ is the set of all possible parts $\wp_i$ defining bipartitions for the state. Consider, for instance, a 4-partite state $\ket{\psi}\in\mc{H}_1\otimes \mc{H}_2\otimes \mc{H}_3\otimes \mc{H}_4$. In this case, $\wp_i$ may assume the labels $\{1, 2, 3, 4, 12, 13, 14\}$,
corresponding to the bipartitions $1|234$, $2|134$, $3|124$, $4|123$, $12|34$, $13|24$, and $14|23$. The rationale behind measure \eqref{En} is as follows. The occurrence of a single factorizable bipartition is enough to make GME null, whereas one has $\mbb{E}_n>0$ iff all reduced states $\rho_{\wp_i}$ are mixed, meaning that biseparability is not admissible, no matter which bipartition is regarded. Because $0\leqslant\mc{L}(\rho)\leqslant 1-\frac{1}{d}$, with $d$ the dimension of the space on which $\rho$ acts, it follows that GME saturates to $\sqrt{2(d-1)/d}$, which is only realized for the maximally entangled $n$-partite state $\frac{1}{\sqrt{d}}\sum_{i=1}^d\ket{i}^{\otimes n}$.

%%%%%%%%%%%%%%%%%%%%%%%%%%%%%%
\subsection{Quantum coherence}
\label{sec:QCoh}

Quantum coherence is a quantum resource \cite{costa2020} which lies at the heart of foundational aspects of the quantum realm as matter wave interference \cite{angelo2015} and realism \cite{bill2015,gomes2018,dieguez2018,gomes2019,freire2019,orthey2019,fucci2019,engelbert2020}, with applications ranging from quantum information \cite{marvian2016} to biological systems \cite{engel2007}. Its quantification with respect to an observable $A$'s eigenbasis is provided by the so-called relative entropy of coherence, $\mbb{C}_A(\rho)\coloneqq S\left(\Phi_A(\rho)\right)-S(\rho)$ \cite{QCoh2014}, where $S(\rho)\coloneqq-\Tr\left(\rho\ln{\rho}\right)$ stands for the von Neumann entropy of the single-partite state $\rho$.  $\Phi_A(\rho)\coloneqq\sum_iA_i\rho A_i$ is the dephasing map associated with unrevealed measurements of the observable $A=\sum_ia_iA_i$, with eigenvalues $a_i$ and projectors $A_i=\ket{a_i}\bra{a_i}$ such that $A_iA_j=\delta_{ij}A_i$. Roughly speaking, $\mbb{C}_A$ tells us how far the state $\rho$ under scrutiny is from its ``decohered'' counterpart $\Phi_A(\rho)$. Starting from the decomposition of a generic $\rho$ in the $\{\ket{a_i}\}$ basis, we obtain that $\Phi_A(\rho) = \sum_k\rho_{kk}A_k\equiv \rho_\text{diag}$, which has no off-diagonal terms in the $A$ basis.

Since $S$ and the linear entropy $\mc{L}$ are widely believed to be monotonic functions of each other, here we propose to measure quantum coherence in terms of such ``linear metric,'' so to speak; that is, we introduce 
\begin{equation}
 \mbb{C}_A(\rho)\coloneqq \mc{L}\left(\Phi_A(\rho)\right)-\mc{L}(\rho)=\Tr\left(\rho^2-\rho_\text{diag}^2\right).
\end{equation}
Noticing that $\Tr\big[(\rho-\rho_\text{diag})^2\big]=\Tr\rho^2+\Tr\rho_\text{diag}^2-2\Tr\rho_\text{diag}^2$ and $\Tr\left[\rho\Phi_A(\rho)\right]=\Tr\big(\rho_\text{diag}^2\big)$, the above formula simplifies to
\begin{equation}
\mbb{C}_A(\rho)=||\,\rho-\rho_\tx{diag}\,||^2,
\label{Coh}
\end{equation}
where $||O||\coloneqq \sqrt{\Tr(O^2)}$ is the Hilbert-Schmidt norm of the observable $O$. It is clear that Eq. \eqref{Coh} furnishes a reasonable estimate for coherence since $\mbb{C}_A$ is null only if $\rho=\rho_\tx{diag}$ and increases with the number of off-diagonal terms in $\rho-\rho_\tx{diag}$. Quantum coherence is, by construction, a basis-dependent notion. In this work, we confine our attention to the observables $P_{0,x,z}^\nu$ and $\sigma_z^\nu$, which are the ones directly involved in the state~\eqref{Psi_before}.

%%%%%%%%%%%%%%%%%%%%%%%%%%%%%%%%%%%%%%%%%%%%%%%%%%%%%%%%%%%%%%%%%%
\section{GME and quantum coherence for the electron-positron pair}

We are now equipped with the tools to discuss how GME and quantum coherence, as given by Eqs. \eqref{En} and \eqref{Coh}, behave upon changes of inertial reference frames. We start by computing these resources with respect to the laboratory perspective, wherein the electron-position state \eqref{Psi_before} is assumed to have been prepared.

%%%%%%%%%%%%%%%%%%%%%%%%%%%%%%%%%%%%%%%%%%%%%%%
\subsection{Quantum resources in the laboratory frame}

The evaluation of GME for the 8-partite state $\Psi=\ket{\Psi}\bra{\Psi}$ given by Eq. \eqref{Psi_before} is trivial because the parts $\mc{P}_{0,x}^\nu$ (corresponding to the $p_{0,x}$ component of the system $\nu$) factorize, so that $\mbb{E}_8(\Psi)=0$. Likewise, because every reduced state $\rho_{\wp_i}$, for $\wp_i\in\{\mc{P}_0^\nu,\mc{P}_x^\nu,\mc{P}_z^\nu,\mc{S}^\nu\}$, is a two-branch mixture (a diagonal state) with probabilities $\eta^2$ and $\xi^2$, we have $\mbb{C}_{A}(\rho_{\wp_i})=0$ for $A\in\{P_0^\nu,P_x^\nu, P_z^\nu,\sigma_z^\nu\}$. However, all this does not imply that the state \eqref{Psi_before} is devoid of quantum resources. In fact, as we show now, genuine $4$-partite entanglement is present. 

Tracing the four parts $\mc{P}_{0,x}^\nu$ out of the global state $\Psi$ gives another pure state $\rho=\Tr_{\mc{P}_{0,x}^\nu}(\Psi)$. According to the prescription \eqref{En}, to compute genuine $4$-partite entanglement, we have to evaluate the bipartite entanglement of the following bipartitions: 
$\mc{P}_z^-|\mc{P}_z^+ \mc{S}^- \mc{S}^+$, $\mc{P}_z^+|\mc{P}_z^- \mc{S}^- \mc{S}^+$, $\mc{S}_z^-|\mc{P}_z^- \mc{P}_z^+ \mc{S}^+$, $\mc{S}_z^+|\mc{P}_z^- \mc{P}_z^+ \mc{S}^-$, $\mc{P}_z^-\mc{P}_z^+ |\mc{S}^- \mc{S}^+$, $\mc{P}_z^-\mc{S}^-|\mc{P}_z^+ \mc{S}^+$, and $\mc{P}_z^-\mc{S}^+|\mc{P}_z^+ \mc{S}^- $. For the first bipartition, for example, we derive the reduced density operator $\rho_{\mc{P}_z^-} = \Tr_{\mc{P}_z^+ \mc{S}^\nu}(\rho) = \eta^2 \ket{\mf{p}}\bra{\mf{p}} + \xi^2 \ket{-\mf{p}}\bra{-\mf{p}}$, from which we obtain the bipartite entanglement $\mc{L}(\rho_{\mc{P}_z^-})=\frac{1}{2}\sin^2(2\phi)$. The calculation is similar for the other bipartitions and the resulting entropies are all the same. Hence, from formula \eqref{En}, we obtain
\begin{equation}
\mbb{E}_4(\Psi)=\sin(2\phi),
\label{E4}
\end{equation}
which turns out to be the only resource available in the laboratory frame (among the ones considered in this work, namely, pure-state entanglement and quantum coherence).

%%%%%%%%%%%%%%%%%%%%%%%%%%%%%%%%%%%%%%%%%%%%%%%%%%%
\subsection{Quantum resources in the boosted frame}

We now investigate how the quantum resources change if one applies a Lorentz transformation to the state (\ref{Psi_before}). Direct application of the formula \eqref{Uket} yields $\ket{\Psi'}=U(\Lambda)\ket{\Psi}$, where
\begin{align}
\ket{\Psi'}&=\eta\ket{\pi_0^+,\pi_x^+,\pi_z^+,u^+}\ket{\pi_0^-,\pi_x^-,\pi_z^-,u^-}\nonumber \\ &+ \xi e^{i\theta}\ket{\pi_0^-,\pi_x^-,\pi_z^-,v^-}\ket{\pi_0^+,\pi_x^+,\pi_z^+,v^+}
\end{align}
and
\begin{equation}
\begin{array}{rcl}
\ket{u^\mf{s}} &=& \cos\left(\frac{\Omega}{2}\right)\,\ket{0}+\mf{s}\sin\left(\frac{\Omega}{2}\right)\,\ket{1},\\ \\
\ket{v^\mf{s}} &=&-\mf{s}\sin\left(\frac{\Omega}{2}\right)\,\ket{0}+\cos\left(\frac{\Omega}{2}\right)\,\ket{1}.
\end{array}
\end{equation}

Let us now compute the GME for the 8-partite boosted state $\Psi'=\ket{\Psi'}\bra{\Psi'}$. We start by looking at the eight bipartitions that separate one part from the other seven, hereafter called the $1|7$ bipartitions. Since $\braket{\pi_a^+|\pi_a^-}=0$ ($a\in\{0,x,z\}$), the reduced states for the parts $\mc{P}_a^\nu$ are statistical mixtures given by  $\rho_{\mc{P}_a^\nu}'=\eta^2\ket{\pi_a^+}\bra{\pi_a^+}+\xi^2\ket{\pi_a^-}\bra{\pi_a^-}$. Hence, for these six reduced states, we find $\mc{L}(\rho_{\mc{P}_a^\nu}')=\frac{1}{2}\sin^2(2\phi)$. For the two remaining $1|7$ bipartitions, the ones referring to the spin parts, the situation is different because the new spin states are not orthogonal, that is, $\braket{u^+|v^-}=\braket{u^-|v^+}=\sin\Omega$. Thus, while the reduced states have the form $\rho_{\mc{S}^{\mp}}'=\eta^2\ket{u^\pm}\bra{u^\pm}+\xi^2\ket{v^\mp}\bra{v^\mp}$, the linear entropy results in $\mc{L}(\rho_{\mc{S}^\nu}')=\frac{1}{2}\sin^2(2\phi)\cos^2(\Omega)$. For almost all the $\frac{8!}{2!6!}=28$ bipartitions of the form $2|6$, involving the orthonormal basis such as $\{\ket{\pi_a^+,u^+},\ket{\pi_a^-,v^-}\}$, we find the same bipartite entanglement $\frac{1}{2}\sin^2(2\phi)$. The only special $2|6$ bipartition is the one referring to the reduced state $\rho_{\mc{S}^-\mc{S}^+}'=\eta^2\ket{u^+,u^-}\bra{u^+,u^-}+\xi^2\ket{v^-,v^+}\bra{v^-,v^+}$, for which one has $\mc{L}(\rho_{\mc{S}^-\mc{S}^+}')=\frac{1}{2}\sin^2(2\phi)(1-\sin^4\Omega)$. The number of $3|5$ bipartitions is $\frac{8!}{3!5!}=56$. In this scenario, there is no special case since all the pertinent bases are orthonormal, even $\{\ket{\pi_a^+,u^+,u^-},\ket{\pi_a^-,v^-,v^+}\}$. The bipartite entanglement is, once again, $\tfrac{1}{2}\sin^2(2\phi)$. Finally, we look at the $\frac{1}{2}\frac{8!}{4!4!}=35$ bipartitions of the form $4|4$, for which the bipartite entanglement is no different from the previous one. To obtain the GME prescribed by Eq. \eqref{En}, we take the minimum value of $\mc{L}$ among the ones indicated in this paragraph, so as to obtain
\begin{equation} 
\mbb{E}_8(\Psi')=\mbb{E}_4(\Psi)\,\cos{\Omega},
\label{E8l}
\end{equation}
with $\cos\Omega$ always being non-negative for $\alpha\in\big(0,\frac{\pi}{2}\big)$. Naturally, one may wonder what happens with $\mbb{E}_4$ in the boosted frame. The first point to note is that we cannot obtain a 4-partite pure state out of $\Psi'$ so that even if $\mbb{E}_4(\rho_{\wp_i}')$ existed, it would not rigorously have the same character as $\mbb{E}_4(\Psi)$. It turns out, though, that $\mbb{E}_4(\rho_{\wp_i}')=0$ for all possible 4-partite configurations $\wp_i$, the reason being the fact that the possible reduced states $\rho_{\wp_i}'$ are all explicitly multiseparable due to the orthogonality of the momenta bases. That is, these states admit the form $\rho_{\wp_i}=\sum_kc_k \rho_{1k}\otimes\rho_{2k}\otimes\rho_{3k}\otimes\rho_{4k}$, which implies full separability and hence no entanglement.

To compute quantum coherence, we return to the $1|7$ bipartitions. Being incoherent mixtures, the reduced states $\rho_{\mc{P}_a^\nu}'$ are already diagonal in the momenta bases, so that $\mbb{C}_{P_a^i}(\rho_{\mc{P}_a^\nu}')=0$ for $a\in\{0,x,z\}$. On the other hand, for the spin reduced states, one shows that $\rho_{\mc{S}^\nu}'-(\rho_{\mc{S}^\nu}')_\tx{diag}=-\nu\frac{\sin{\Omega}}{2}(\ket{0}\bra{1}+\ket{1}\bra{0})$ for $\nu\in\{-,+\}$, and then, via formula \eqref{Coh}, we obtain 
\begin{equation}
\mbb{C}_{\sigma_z^\nu}(\rho_{\mc{S}^\nu}')=\frac{1}{2}\sin^2\Omega.
\label{Csigmazl}
\end{equation}
%

%%%%%%%%%%%%%%%%%%%%%%%
\subsection{Discussion}

The results expressed by Eqs. \eqref{E4}, \eqref{E8l}, and \eqref{Csigmazl} are formally simple and insightful. First, they do not depend on the relative phase $\theta$, a fact that is well known from the entanglement theory. Second, the quantum resources in the boosted frame are clearly modulated by the factor $\cos{\Omega}$ [Eq.~\eqref{cos-Omega}], which emerged from the Wigner rotation and hence critically depends on the boost parameters.

\begin{figure}[htb]
\centering
\includegraphics[width=\columnwidth]{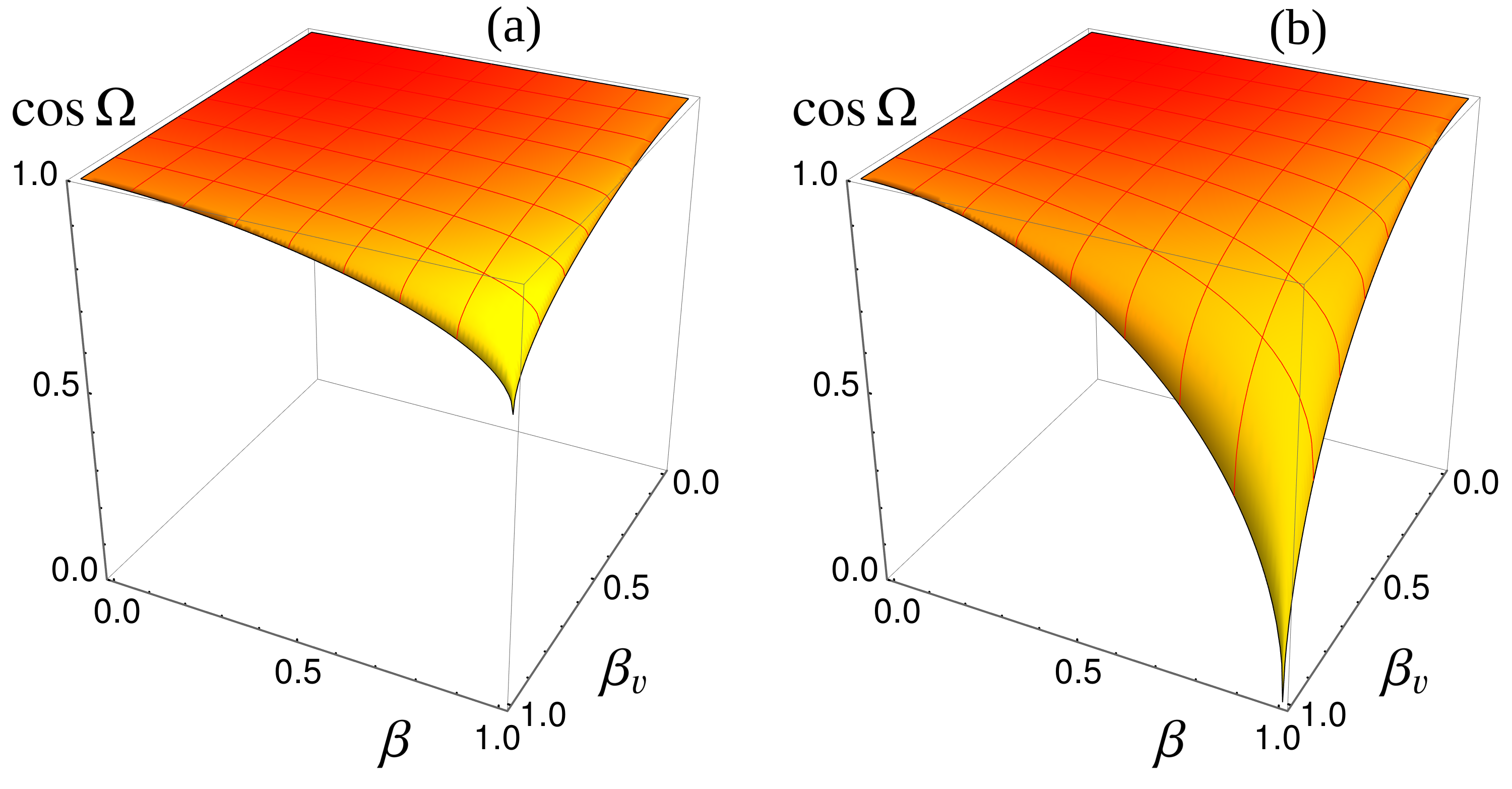}
\caption{\small Transformation factor $\cos{\Omega}=\mbb{E}_8(\Psi')/\mbb{E}_4(\Psi)$ for GME as a function of $\beta$ (dimensionless velocity of the boost) and $\beta_v$ (dimensionless velocity of the particles), for (a) $\alpha=\pi/4$ and (b) $\alpha=\pi/2$. The entanglement attenuation becomes appreciable only in the high-velocity regime. In particular, in the ultrarelativistic limit, one has $\lim_{\beta\to1}\lim_{\beta_v\to 1}\cos{\Omega}=\cos\alpha=\hat{\beta}\cdot\hat{\beta}_v$, where $\hat{\beta}_v=\hat{p}=\vec{p}/||\vec{p}||$.}
\label{fig2}
\end{figure}

As far as pure-state entanglement is concerned, we have an interesting and nontrivial interchange in the structure of this resource: genuine 4-partite entanglement in the laboratory frame is transformed into genuine 8-partite entanglement (and coherence). This means that, strictly speaking, we cannot claim entanglement invariance, not even when $\cos{\Omega}\to 1$ (which occurs as $\alpha\to 0$) because, in the boosted frame, entanglement is codified in a larger number of degrees of freedom, thus being different in essence. Figure \eqref{fig2} illustrates how $\mbb{E}_8(\Psi')$ gets attenuate in relation to $\mbb{E}_4(\Psi)$ as a function of the dimensionless velocities $\beta$ and $\beta_v$. The attenuation becomes significant (and abrupt) in the high-velocity limit. In particular, it follows from this analysis that the GME in the boosted frame is lower bounded as $\mbb{E}_8(\Psi')\geqslant \mbb{E}_4(\Psi)\,\cos{\alpha}$, with equality holding only in the ultrarelativistic limit  $(\beta,\beta_v)\to (1,1)$. 

As expected, the amount of entanglement is controlled by the term $2\eta\xi=\sin(2\phi)$, for $\eta$ and $\xi$ determine the characteristics of the quantum superposition in the electron-positron state \eqref{Psi_before}. Accordingly, no quantum resources whatsoever are available in the laboratory frame when $\phi\in\big\{0,\frac{\pi}{2}\big\}$. Still, quantum coherence will manifest itself in the boosted frame because the relative motion induces a rotation---a Wigner rotation, $\mf{D}_{\mf{s}\hat{y}}(\Omega)$---in the axis along which the spin is quantized, thus leading the spin to enter in quantum superposition. Moreover, the quantum coherence associated with the $z$ component of spin increases as $\mbb{E}_8(\Psi')$ decreases, reaching its maximum in the ultrarelativistic regime. In the regime where both $\beta$ and $\beta_v$ are small enough, one has $\mbb{C}_{\sigma_z^\nu}(\rho_{\mc{S}^\nu}')\cong \frac{1}{8}\beta^2\beta_v^2\sin^2\alpha\ll 1$, showing that coherence invariance can be claimed to some extent. 

%%%%%%%%%%%%%%%%%%%%%%%%%%%%%%%%%%%%%%%%%%%%%
\subsection{Constructing a quantum invariant}

For simplicity, let us introduce the notation $\mbb{E}_4\equiv\mbb{E}_4(\Psi)$, $\mbb{E}_8\equiv\mbb{E}_8(\Psi)$, $\mbb{C}_{\nu}\equiv\mbb{C}_{\sigma_z^\nu}(\rho_{\mc{S}^\nu})$, and, similarly for the boosted frame, $\mbb{E}_4'\equiv\mbb{E}_4(\rho_{\wp_i}')$, $\mbb{E}_8'\equiv\mbb{E}_8(\Psi')$, $\mbb{C}_{\nu}'\equiv\mbb{C}_{\sigma_z^\nu}(\rho_{\mc{S}^\nu}')$. Via direct algebra, we can combine the results of the previous section to arrive at $\mbb{E}_4=\mbb{E}_8'/\sqrt{1-(\mbb{C}_+'+\mbb{C}_-')}$. Now, since $\mbb{C}_\pm=\mbb{E}_8=0$, we can compose the invariant
\begin{equation}
\frac{\mbb{E}_4+\mbb{E}_8}{\sqrt{1-(\mbb{C}_++\mbb{C}_-)}}=\frac{\mbb{E}_4'+\mbb{E}_8'}{\sqrt{1-(\mbb{C}_+'+\mbb{C}_-')}}=\sin(2\phi),
\label{invariant}
\end{equation}
which is maximized for a Bell-like state $(\phi=\pi/4)$.
Although one may claim that this quantity emerged artificially, it offers some hint of what kind of information-theoretic object should be taken into account for one to obtain the definitive quantum invariant, conceivable by first principles of information theory. That such a quantum invariant must exist is an intuitive idea grounded in the fact that the total information encoded in a quantum state does not change upon unitary transformations~\cite{savi2021}, including Lorentz boosts. That the invariant \eqref{invariant} is not expected to be the definitive one is suggested by the fact we have considered neither other types of entanglement, such as mixed-state entanglement, nor quantum coherences associated with observables other than $\sigma_z$.

%

%%%%%%%%%%%%%%%%%
\section{Summary}

In this work, we investigated how some quantum resources, such as genuine multipartite entanglement and quantum coherence, transform upon Lorentz boosts. To this end, we considered (i) an electron-positron multipartite pure state, (ii) Lorentz boosts perpendicular to the Wigner rotation axis, and (iii) measures of GME and quantum coherence based on the linear entropy. We found an interesting state of affairs in which genuine 4-partite entanglement in the laboratory frame transforms to genuine 8-partite entanglement plus quantum coherences in the $\sigma_z^\nu$ bases. Furthermore, within the adopted framework, we showed that a given combination of quantum resources can be derived which remains invariant upon the Lorentz boosts. 

It is noteworthy that our analysis was not aimed at exhausting the whole universe of quantum resources (see Ref.~\cite{costa2020} for a small collection of them). In fact, even other entanglement structures (such as $\mbb{E}_{3\leqslant n<8}$ for mixed reduced states) and quantum coherence (for instance, $\mbb{C}_{\sigma_{x,y}^\nu}$) were set aside. This constitutes a fruitful research program for near future works.
  
%======================
\begin{acknowledgments}
The authors acknowledge Nicholas G. Engelbert for useful discussions. A.T.P. acknowledges financial support from an undergraduate scholarship from UFPR Tesouro Nacional. This research was supported in part by Grants No. NSF DMR-1606591 (G.C.) and No. CNPq/Brazil 309373/2020-4 (R.M.A). Support from the National Institute for Science and Technology of Quantum Information (INCT-IQ/CNPq, Brazil) also is gratefully acknowledged (R.M.A.).
\end{acknowledgments}
%======================

\bibliography{Refs}

\end{document}